\documentstyle[preprint,tighten,aps,floats,psfig,epsfig,latexsym]{revtex}

\def\szed{\hbox{\rm Z\kern-0.45em\hbox{\rm Z}}}  

\begin{document}
\draft
\title{ 
The normal-to-planar superfluid transition in $^3$He
}
\author{Martino De Prato,${}^{a}$ 
        Andrea Pelissetto,${}^{b}$
        Ettore Vicari${}^{c}$
}
\address{${}^a$ Dipartimento di Fisica, Universit\`a di Roma Tre, and 
              INFN, I-00146 Roma, Italy} 
\address{${}^b$ Dipartimento di Fisica, Universit\`a di Roma La Sapienza,
and INFN, I-00185 Roma, Italy} 
\address{${}^c$ Dipartimento di Fisica, Universit\`a di Pisa, and 
              INFN,
              I-56127 Pisa, Italy \\ 
{\bf e-mail: \rm 
{\tt deprato@fis.uniroma3.it},
{\tt Andrea.Pelissetto@roma1.infn.it},
{\tt vicari@df.unipi.it}.
}}

\date{\today}

\maketitle

\begin{abstract}
We study the nature of the $^3$He superfluid transition from the normal to the 
planar phase, which is expected to be stabilized by the dipolar interactions.
We determine the RG flow of the corresponding Landau-Ginzburg-Wilson theory
by exploiting  two fixed-dimension perturbative schemes: the massive 
zero-momentum scheme and the minimal-subtraction scheme without $\epsilon$ 
expansion. The analysis of the corresponding six-loop and five-loop series 
shows the presence of a stable fixed point in the relevant coupling region. 
Therefore, we predict the transition to be continuous. We also compute critical 
exponents. The specific-heat exponent $\alpha$ is estimated as 
$\alpha = 0.20(15)$, while the magnetic susceptibility and magnetization exponents 
$\gamma_H$ and $\beta_H$ for $^3$He are 
$\gamma_H = -0.34(5)$, $\beta_H = 1.07(9)$. 
\end{abstract}


\section{Introduction and summary}
\label{intro}

The superfluid phases in $^3$He have been extensively studied in the 
years \cite{Leggett-75,Wheatley-75,VW-90,Lee-97}. In the absence of 
magnetic field two superfluid phases, named A and B respectively, have 
been identified. However, if one takes into account the 
weak dipolar interaction \cite{Leggett-74,Leggett-75}, a new phase, 
called planar phase, should separate the normal state from the 
A and B superfluid states \cite{JLM-76,BLM-77}. At present, 
there is no experimental evidence for a normal-to-planar transition.
This is due to the fact that dipolar interactions 
are expected to become relevant only within the interval
$|t|\lesssim 2 \times 10^{-6}$, where $t\equiv 1 - T/T_c$. 
For pressures between 10 and 30 bar, $T_c$ varies between 2 and 2.6 mK, 
so that the interesting temperature interval is approximately
$|T-T_c|\approx$ (4-5)$\cdot 10^{-9}$ K. Outside this interval 
normal Fermi liquid or A/B superfluid behavior should be observed.
Despite the narrowness of such temperature
interval, Ref.~\cite{JLM-76} argued that it should be 
possible to observe  the planar phase if the temperature $T$ can be 
controlled to high precision, as it has been done recently for 
$^4$He \cite{He4}.
It is therefore interesting to investigate its nature in detail.

The normal-to-superfluid transition in $^3$He
is supposed to be driven by $p$-wave spin-triplet pairing states.
Near the transition the critical behavior is controlled by an order
parameter that is  a complex 3$\times$3 matrix, see, e.g., 
Refs.~\cite{MS-73,BA-73,BM-74,Leggett-75}. The weak dipolar interaction
breaks the original SO(3)$\times$SO(3)$\times$U(1) symmetry down to 
SO(3)$\times$U(1) and the corresponding order parameter becomes a 
complex three-component field $\psi$. If we 
neglect strain gradient effects \cite{AGR-74}, 
which are expected to be irrelevant at the transition \cite{JLM-76},
the Landau-Ginzburg-Wilson Hamiltonian describing the critical modes 
at the normal-to-planar transition can be written as
\cite{JLM-76,BLM-77,footnoteH} 
\begin{equation}
{\cal H} = \int d^d x
\left[  
\partial_\mu \psi^*\cdot \partial_\mu \psi + r \psi^* \cdot \psi  
+{1\over 4} 
f_0 \left( \psi^*\cdot \psi\right)^2 + 
{1\over 4}  g_0 |\psi\cdot \psi |^2 \right],
\label{LGWpsi}
\end{equation}
with $f_0>0$ and $g_0<0$. Approximate estimates 
\cite{JLM-76,AB-73,BSA-74,footnote1}
give $g_0/f_0\approx -1/3$. 
Note that stability requires $f_0\geq 0$ and $f_0+g_0\geq 0$.
When $g_0<0$ the Hamiltonian is minimized by fields 
$\psi_a = e^{i\theta} n_a$, where $n_a$ is a constant real vector, and the
corresponding ground-state manifold is $(S^2\times S^1)/\szed_2$,
where $S^n$ is the $n$-dimensional sphere. The case $g_0>0$ with a 
two-component and a three-component field $\psi$  is also physically
interesting because it describes the critical
properties of frustrated spin models with
noncollinear order \cite{Kawamura-88,Kawamura-98}.
The same Hamiltonian has also been considered to discuss the 
critical properties of Mott insulators \cite{Sachdev-02}.
It should also be relevant for 
two-dimensional quantum phase transitions in 
cuprate superconductors \cite{ZDS-02}
and for transitions in heavy-fermion compounds such as UPt$_3$ \cite{JT-02}.

In the mean-field approximation the model with Hamiltonian (\ref{LGWpsi})
has a continuous second-order transition.
On the other hand, $\epsilon$-expansion calculations  
for an $N$-component field $\psi$
find \cite{JLM-76,BLM-77,Kawamura-88,ASV-95} a stable fixed point (FP) 
in the region $g_0<0$ only for 
$2 - \epsilon \lesssim N \lesssim 2.20 - 0.57\epsilon$, $\epsilon \equiv 4 - d$.
Apparently no FP is found for $N=3$,
suggesting a fluctuation-induced first-order transition.
This scenario has been further supported by three-loop calculations
within a three-dimensional (3-$d$) massive zero-momentum (MZM) scheme 
\cite{AS-94} and by a nonperturbative renormalization-group (RG) study 
of the effective average action in the lowest-order 
approximation of the derivative expansion \cite{KW-01}.
As we shall argue below mentioning a few specific physical examples, 
these results may not be conclusive. 
For instance, in some physically interesting cases, 
low-order, and in some cases also high-order,
$\epsilon$-expansion calculations  fail to provide
the correct physical picture.  The location and the stability of
the FP's may drastically change approaching $d=3$,
and new FP's, not present for $d\approx 4$, may appear in three dimensions.  
We mention the Ginzburg-Landau model of superconductors, in which 
a complex scalar field couples to a gauge field.
One-loop $\epsilon$-expansion  calculations \cite{HLM-74}
indicate that no stable FP exists unless the number $N$ of real components
of the scalar field is larger than $N_c=365 + O(\epsilon)$.
This number is much larger than the physical value $N=2$.
Consequently, a first-order transition was always expected \cite{HLM-74}.
Later, exploiting 3-$d$ theoretical approaches 
(see, e.g., Ref.~\cite{kleinertbook}) and
Monte Carlo simulations (see, e.g., Ref.~\cite{MCs}),
it was realized that 3-$d$ systems described by the 
Ginzburg-Landau model can also undergo a continuous transition---this
implies the presence of a stable FP in the 3-$d$ Ginzburg-Landau theory---in
agreement with experiments \cite{GN-94}.
A similar phenomenon probably occurs---but this is still a  
controversial issue, see, e.g., Refs.~\cite{Kawamura-98,PV-rev,DMT-03}---in 
model (\ref{LGWpsi}) when $g_0>0$.
$\epsilon$-expansion calculations find a stable FP
with attraction domain in the region $g_0>0$ only for
$N>N_c$ with \cite{Kawamura-88,ASV-95,PRV-01b,CP-03} 
$N_c = 21.80+ O(\epsilon)$,
which seems to exclude the physically interesting cases $N=2,3$.
These conclusions were apparently confirmed by three-loop 3-$d$
perturbative calculations in the MZM scheme \cite{AS-94} and by 
a nonperturbative RG study of the effective average action  \cite{TDM-00}, 
using approximations based
on the lowest orders of the derivative expansion.
On the other hand, recent studies based on 3-$d$ high-order
field-theoretical calculations found a stable FP with attraction domain
in the region $g_0>0$ for both $N=2$ and $N=3$.  Two different
renormalization schemes were used: the MZM scheme
(six loops) \cite{PRV-01,CPS-02} and 
the minimal-subtraction ($\overline{\rm MS}$) scheme without 
$\epsilon$ expansion (five loops) \cite{CPPV-03-inprep}. 
These results confirm the existence of a new chiral universality class 
and are in agreement with most experiments that observe 
continuous transitions in stacked triangular antiferromagnets
\cite{Kawamura-98,PV-rev,CP-97}.

In this paper we investigate the nature of the
normal-to-planar transition in $^3$He by exploiting two different
3-$d$ perturbative field-theoretical approaches  
based on the Hamiltonian (\ref{LGWpsi}).
The first one, the  MZM scheme, is defined in the massive (disordered)
phase \cite{Parisi-80}. It is based on a zero-momentum renormalization
procedure and an expansion in powers of zero-momentum
quartic couplings.
As already mentioned, this perturbative
scheme was considered in Ref.~\cite{AS-94},
where three-loop series were analyzed without finding evidence 
for stable FP's.
The second scheme is the so-called $\overline{\rm MS}$ 
scheme without $\epsilon$ expansion \cite{SD-89}, which is 
defined in the massless critical theory.
This scheme is strictly related to the $\epsilon$ expansion,
but, unlike it, no expansion in powers of $\epsilon$ is performed.
Indeed, $\epsilon$ is set to its physical value $\epsilon=1$, 
and one analyzes the expansions in powers of the 
renormalized quartic couplings.

Here we shall present an analysis of the six-loop MZM
series that have been computed in Ref.~\cite{PRV-01}
and of the five-loop $\overline{\rm MS}$ 
series that have been computed in Ref.~\cite{CP-03}.
We find a stable FP in the region $g<0$ whose attraction domain 
includes the region where $g_0/f_0\approx -{1/3}$, which should be the 
relevant one for the transition in $^3$He \cite{footnote1}.
The two perturbative schemes give consistent results,
providing a nontrivial crosscheck of the results.
We thus predict a continuous normal-to-planar transition,
belonging to a new universality class.
Our result contradicts earlier
theoretical works, see, e.g., Refs.~\cite{JLM-76,AS-94,KW-01}.

We also determine the standard critical exponents for model
(\ref{LGWpsi}). We obtain
$\nu = 0.59(4)$, $\eta = 0.079(7)$, $\gamma = 1.14(8)$, 
$\beta = 0.32(3)$ in the MZM scheme and 
$\nu = 0.63(8)$, $\eta = 0.086(24)$, $\gamma = 1.20(15)$, 
$\beta = 0.34(4)$ in the $\overline{\rm MS}$
scheme. A weighted average 
gives
\begin{equation}
\nu=0.60(5),\quad  \eta=0.08(1), \quad  \gamma = 1.16(10), \quad
\beta = 0.33(4). 
\end{equation}
The errors are such to include the MZM results 
(with their errors) that apparently are the most 
precise ones.  By using hyperscaling we also obtain $\alpha = 0.20(15)$.
Note that, while $\nu$ and $\alpha$ are indeed the critical exponents 
associated with the specific heat and with the correlation length, 
$\gamma$, $\eta$, and $\beta$ are not directly accessible in $^3$He and 
are {\em not} related with the magnetic critical behavior of $^3$He.
Indeed, the operator that couples with the external magnetic field 
in $^3$He is the so-called chiral operator 
\cite{Kawamura-88,Kawamura-98,PRV-02}
$C_{a} \equiv i \epsilon_{abc} \psi^*_b \psi_c$.
We compute the exponents $\gamma_H$ and $\beta_H$ that characterize the 
singular behavior of the magnetic susceptibility and of the magnetization, and 
$\eta_H$, related to the large-momentum behavior of the magnetic 
structure factor (in field-theoretical terms it is defined in terms of 
the Fourier transform of 
$\langle C_{a}(0) C_{a}(x)\rangle$ that behaves as $q^{-2+\eta_H}$ 
for large momenta $q$). 
We find
$\eta_H = 2.72(7)$, $\gamma_H = -0.34(5)$, $\beta_H = 1.05(7)$ in the 
MZM scheme and
$\eta_H = 2.63(19)$, $\gamma_H = -0.34(11)$, $\beta_H = 1.11(12)$ 
in the minimal-subtraction scheme. 
A weighted average gives
\begin{equation}
\eta_H=2.70(9), \quad  \gamma_H = -0.34(5), \quad \beta_H = 1.07(9),
\end{equation}   
where the errors have been computed as before. Note that $\gamma_H$ is negative
so that in the high-temperature phase the analytic background gives the 
dominant contribution to the magnetic susceptibility $\chi$.
Close to $T_c$, we have therefore
$\chi \approx {\rm const} + (T_c - T)^{0.34}$. In the low-temperature phase
instead $\chi$ is infinite due to the presence of Goldstone modes.

The paper is organized as follows.
In Sec.~\ref{sec2} we present the analysis of the six-loop series in the 
MZM scheme. It provides a rather robust evidence for the presence
of a stable FP with attraction domain in the region
$g_0<0$, which is the relevant one for the $^3$He superfluid transition.
This result is fully confirmed by the analysis
of the five-loop series in the massless $\overline{\rm MS}$
scheme that is presented in Sec.~\ref{sec3}.
We report the perturbative expansions analyzed here
in the MZM and $\overline{\rm MS}$ schemes 
in App.~\ref{appgexp} and \ref{appmsb}, respectively.

\section{The massive zero-momentum scheme}
\label{sec2}

In this Section we consider the fixed-dimension 
massive zero-momentum (MZM)
scheme that describes the disordered massive
phase. The relevant RG functions were computed to 
six loops in Refs.~\cite{PRV-01,PRV-02}.
In App.~\ref{appgexp} we report the series for the relevant case $N=3$.

The perturbative expansions are asymptotic. Therefore, perturbative series
must be resummed  in order to study the RG flow. 
Consider a generic quantity  $S(f,g)= \sum_{ij} c_{ij} f^i g^j$.
In order to determine $S(f,g)$, one may resum the series 
\begin{equation}
S(x f,x g) = \sum_k s_k(f,g) x^k
\label{seriesx}
\end{equation}
in powers of $x$ and then evaluate it at $x=1$.
This can be done by using the conformal-mapping method \cite{LZ-77,ZJbook},
which exploits the knowledge of the large-order behavior
of the expansion, or the so-called Pad\'e-Borel method.  
The large-order behavior of the coefficients is generally given by
\begin{equation}
s_k(f,g) \sim k! \,[-A(f,g)]^{k}\,k^b\,\left[ 1 + O(k^{-1})\right].
\label{lobh}
\end{equation}
Using semiclassical arguments, one can argue that \cite{PRV-01}
the expansion is Borel summable for
\begin{equation}
f \geq 0,\qquad f + g \geq 0.
\label{brr}
\end{equation}
In  this domain we have
\begin{equation}
A(f,g) = a \,{\rm Max} \left[ f,f+g\right] > 0,
\label{afg}
\end{equation}
where $a\approx 0.14777422$.
Under the additional assumption that all the singularities of the 
Borel transform lie on the negative axis,
the conformal-mapping method turns the original
expansion into a convergent one \cite{LZ-77} when 
conditions (\ref{brr}) are satisfied.
Outside this region the expansion is not Borel summable.
However, if the condition
\begin{equation}
f + \case{1}{2} g > 0
\label{brr2}
\end{equation}
holds, then
the singularity of the Borel transform that is closest to the origin
is still in the negative axis.
Therefore, the leading large-order behavior is still
given by (\ref{lobh}) with $A(f,g)$ given by (\ref{afg}).
In this case, the conformal-mapping method 
is still able to take into account the leading large-order behavior,
although it does not provide a convergent series.
Therefore, one may hope to get an asymptotic 
expansion with a milder behavior, which still provides
reliable results.

The RG flow of the theory is determined by the fixed-points (FP's).
Two FP's are easily identified:
the Gaussian FP ($f^*=g^*=0$) and the O(6) FP at \cite{AS-95}
$f^*\approx 0.86$, $g^* = 0$.
The results of Ref.~\cite{CPV-03} on the stability of the 
three-dimensional O($M$)-symmetric FP's
under generic perturbations can be used to prove 
that the O(6) FP is unstable. Indeed, the perturbation term  
$|\psi^2|^2$ in Hamiltonian (\ref{LGWpsi})
is a particular combination of quartic operators transforming as
the spin-0 and spin-4 representations of the O(6) group,
and any spin-4 quartic perturbation is relevant
at the O($M$) FP for $M\geq 3$ \cite{CPV-03},
since its RG dimension $y_{4,4}$ is positive for $M\geq 3$.
In particular, 
$y_{4,4}\approx 0.27$ at the O(6) FP \cite{CPV-03}.

The analyses of the six-loop series reported in
Refs.~\cite{PRV-01,CPS-02} provided a rather robust evidence of the
presence of a stable FP with $g^*>0$ and attraction domain 
in the region $g_0>0$, at $f^*=-0.4(2)$ and $g^*=1.5(1)$.
As already mentioned in the introduction, 
this FP should descrive continuous transitions in 
frustrated three-component spin models with noncollinear order.
However, it is of no relevance for the critical behavior of ${}^3$He. 
Indeed, due to the presence of the unstable O(6) FP,  the $g=0$ axis
plays the role of a separatrix, and thus the RG flow 
corresponding to $g_0<0$ cannot cross the $g=0$ axis.
Therefore, the RG flow relevant for the normal-to-planar
transition is determined by FP's in the region $g<0$.
We shall now show that the RG flow obtained by resumming the six-loop 
$\beta$-functions provides
a rather robust evidence for another stable FP in the region $g<0$ with
attraction domain including the region $g_0/f_0\approx -1/3$,
which should be the relevant one for the $^3$He superfluid transition.

\begin{figure}[tb]
\centerline{\psfig{width=12truecm,angle=-90,file=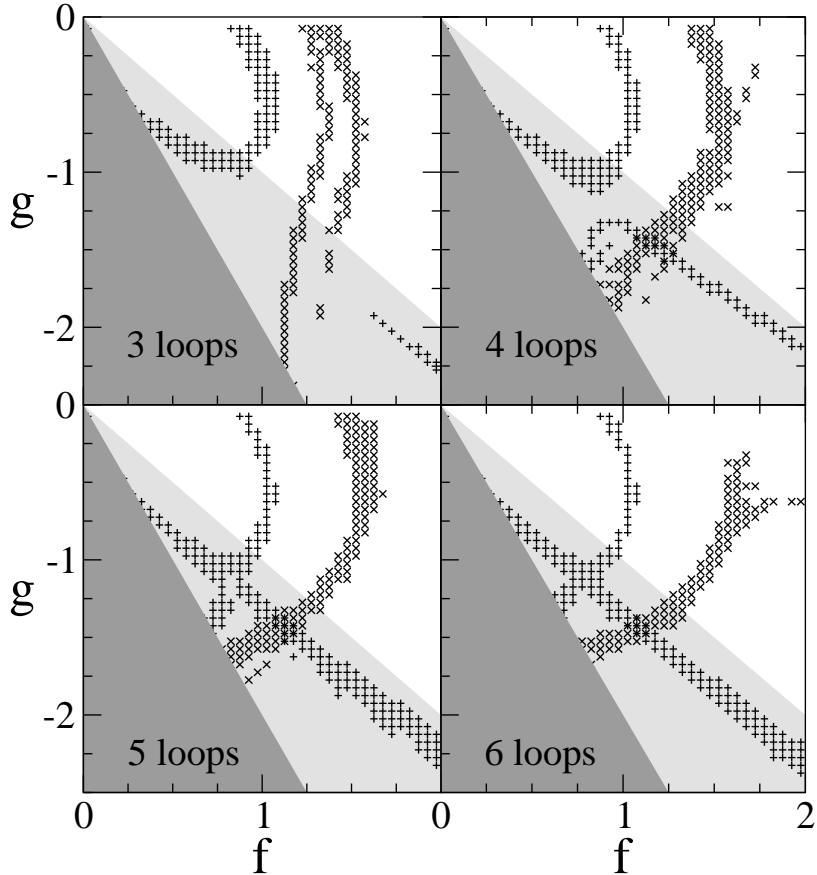}}
\vspace{2mm}
\caption{
Zeroes of the $\beta$-functions in the $f$-$g$ plane in the MZM scheme.
Pluses and crosses correspond to zeroes
of $\beta_f$ and $\beta_g$, respectively.
The FP coordinates are: $f^* = 1.08(12)$, $g^* = -1.46(10)$.
Different colors are used to mark the regions: (1) $f + g > 0$, 
(2) $f + g < 0$ and $f + {1\over2} g > 0$, (3) $f + {1\over2} g < 0$.
}
\label{zeroesMZM}
\end{figure}

In order to investigate the RG flow in the region $g_0<0$, we apply
the same analysis of  Refs.~\cite{PRV-01,CPV-00}.
We refer to these references for details.
We resum the perturbative series by means of 
the conformal-mapping method \cite{LZ-77} that takes into
account the large-order behavior of the perturbative series.
To understand the systematic errors,
we vary two different parameters, $b$ and $\alpha$ (see 
Refs.~\cite{PRV-01,CPV-00} for definitions), in the analysis.
We also apply this method for those values of $f$ and $g$ for which the
series are not Borel summable but still satisfy $f+g/2>0$. 
As already discussed, although in this case the sequence of 
approximations given by the conformal-mapping method is only
asymptotic, it should provide reasonable estimates,
since we are taking into account the leading large-order behavior. 
In Fig.~\ref{zeroesMZM} we report our results for the 
zeroes of the $\beta$-functions, obtained from the analysis of the $l$-loop
series, $l=3,4,5,6$. 
For each $\beta$-function we consider 18 different approximants of 
$\beta(f,g)/f$ 
with $b=3,6,\ldots,18$ and $\alpha=0,2,4$ and we determine the lines
in the $(f,g)$ plane on which they vanish. Then, we divide 
the domain 
$0\le f \le 2$ and $-2.5\le g \le 0$ into 
$40\times50$ rectangles, marking 
those in which at least {\em three} approximants of each $\beta$ function
vanish. No evidence for a FP appears at
three loops, consistently with Ref. \cite{AS-94}.
As the number $l$ of loops increases, a new stable FP---quite stable 
with respect to $l$---clearly appears. 
Such zeroes appear in  9\%, 22\%, 59\%, 89\% 
of the cases we consider 
for $l=3,4,5,6$. Clearly, the new zero is increasingly stable
as $l$ increases. 
At 6 loops we obtain a robust evidence for a FP at
\begin{equation}
f^* = 1.08(12), \qquad g^* = -1.46(10), 
\label{stfp}
\end{equation}
where the error bars correspond to 2 standard deviations.
All zeros of the approximants with $3\le b \le 18$ and $2\le \alpha\le 4$
lie within half of the reported confidence interval and only a few with 
$\alpha = 0$ give significantly different estimates for 
$f^*$ and $g^*$. Moreover, for this smaller class of approximants
a stable FP appears in 100\% (73\%) of the cases for $l=6$ ($l=5$).
Estimate (\ref{stfp}) is stable with respect to the number of loops.
For $l=5$ we obtain $f^* = 1.12(18)$, $g^* = -1.47(18)$. 
Notice that the FP belongs to the region in which the series are not Borel 
summable, but the FP values still satisfy $f^*+g^*/2>0$. 
Therefore, our resummations should be reliable: the stability of the results 
with respect to $l$ supports this expectation.
The FP (\ref{stfp}) is stable, since the 
eigenvalues of the corresponding stability matrix have
positive real part. Our estimate is $\omega_{\pm}=1.03(29+15)\pm i 0.72(11+10)$,
where the first error takes into account the variation of the estimate 
with $\alpha$ and $b$, while the second one takes into account the error
on $f^*$ and $g^*$. Note that the eigenvalues are apparently complex,
suggesting that the FP is actually a focus,
as it happens for the chiral FP \cite{CPS-02}.
Using the RG functions $\eta_\psi$ and $\eta_t$, we have generated 
perturbative series with FP value corresponding to $\nu$, $\eta$, 
$\gamma$, and $\beta$. 
Correspondingly, we obtain the following estimates of the critical exponents:
\begin{eqnarray}
&& \nu=0.59(3+1), \quad \eta=0.079(2+5),\quad
\gamma=1.14(6+2),\quad \beta=0.32(2+1),
\label{exp-MZM}
\end{eqnarray}
where, as before, the first error gives the dependence on $\alpha$ and $b$
and the second is related to the uncertainty on the FP coordinates.
Consistent results are obtained from the analysis of the perturbative series 
for $1/\nu$, $\eta/f^2$, $f^2/\eta$, 
$1/\gamma$, and $1/\beta$. The RG relations among the different critical 
exponents are also well verified.

\begin{figure}[tb]
\centerline{\psfig{width=12truecm,angle=-90,file=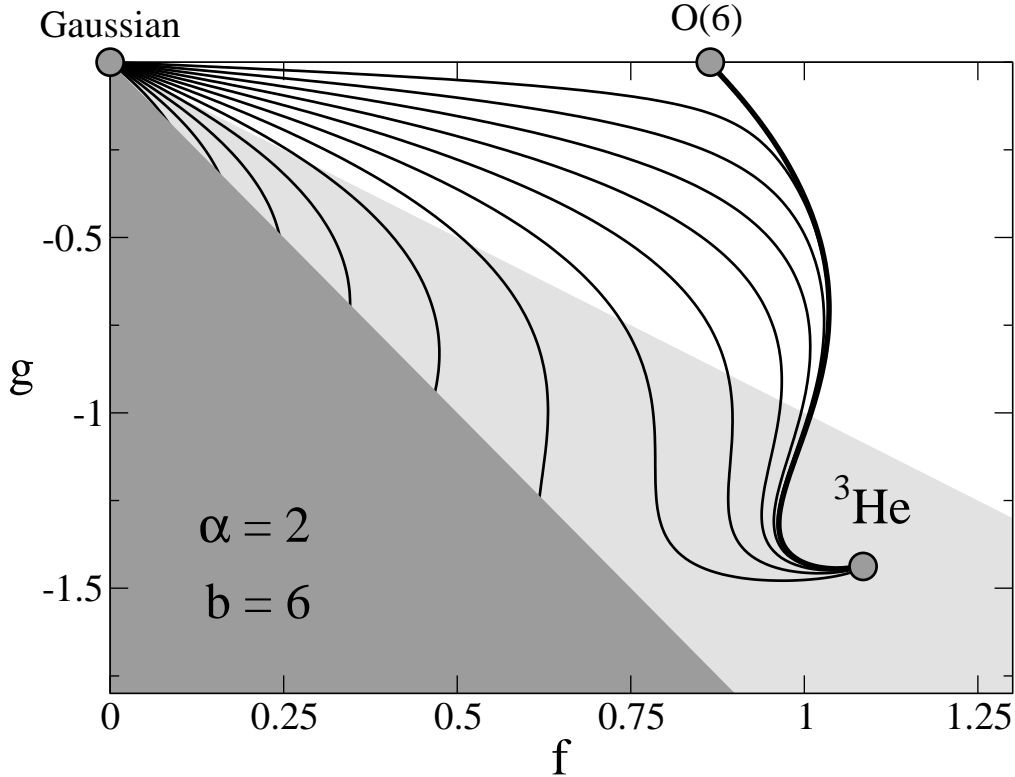}}
\vspace{2mm}
\caption{
RG flow in the $f$-$g$ plane in the MZM scheme. It is obtained by using 
the approximant ($b = 6$, $\alpha = 2$) for $\beta_f(f,g)/f$ and 
$\beta_g(f,g)/f$. 
The thick line is related to the limiting RG trajectory obtained 
for $g_0\rightarrow 0^-$. The other trajectories reaching the FP correspond to 
(from top to bottom)
$g_0/f_0 = -0.085$, $-0.172$, $-0.261$, $-0.354$, $-0.454$, $-0.561$.
Different colors are used to mark the regions: (1) $f + g > 0$, 
(2) $f + g < 0$ and $f + {1\over2} g > 0$, (3) $f + {1\over2} g < 0$.
}
\label{rgflowMZM}
\end{figure}

The exponents reported in (\ref{exp-MZM}) follow the usual nomenclature 
of magnetic systems. However, while $\nu$ is indeed the 
correlation-length exponent in $^3$He, $\eta$, $\gamma$, and $\beta$ 
are not accessible in $^3$He and in particular, are not related to the 
magnetic properties of the superfluid. Indeed, in $^3$He an 
external magnetic field $\vec{H}$ gives rise to a coupling of the 
form \cite{footH} 
\begin{equation}
i\sum_{abc} \epsilon_{abc} H_a \psi_b^* \psi_c. 
\end{equation}
Therefore, the magnetic exponents are obtained by computing 
the RG dimension of the operator 
\begin{equation}
C_{a} = i \epsilon_{abc} \psi^*_b \psi_c
\label{defC}
\end{equation}
Such an operator was discussed at length in 
Refs.~\cite{Kawamura-88,Kawamura-98,PRV-02} in the context of the 
frustrated chiral models and its associated RG dimension was computed at 
the chiral FP with $g>0$. In the present case the analysis of the six-loop
series of Ref.~\cite{PRV-02} at the FP (\ref{stfp}) gives
\begin{equation}
\eta_H=2.72(3+4), \quad \gamma_H=-0.34(2+3) , \quad \beta_H = 1.05(4+3),
\label{chexp-MZM}
\end{equation}
where the two errors are related to the dependence on $\alpha$ and $b$
and to the uncertainty on the FP coordinates.
We have also resummed the series corresponding to $1/\eta_H$, 
$1/\beta_H$, and $1/\gamma_H$. In the first two cases, the results are 
in perfect agreement with the estimates (\ref{chexp-MZM}); only for $\gamma_H$ 
do we find a significant difference. Note, however, that scaling relations 
give $\gamma_H = \nu (2 - \eta_H) =  -0.42 (5)$ 
and  $\gamma_H = 3 \nu - 2 \beta_H = -0.33 (18)$, 
and thus confirm the direct estimate (\ref{chexp-MZM}).

In Fig.~\ref{rgflowMZM} we show the RG flow in the $(f,g)$ plane
for several values of the Hamiltonian parameters $f_0$ and $g_0$ and for a
given approximant of the two $\beta$ functions
(other approximants give similar results). 
We consider values in the stability region $f_0 > 0$ and 
$f_0 + g_0 > 0$ and restrict the calculation to the region of interest
$g_0 < 0$.  We refer to Ref.\cite{CPPV-03} for details on the determination
of the RG trajectories and their relation with the quartic
parameters $f_0$ and $g_0$ of the Hamiltonian.
All trajectories corresponding to $-0.65\lesssim g_0/f_0 < 0$ 
lie completely in the region $f + {1\over2} g > 0$ in which resummations should
be reliable and are attracted by the stable FP. 
Therefore, the attraction domain of the FP (\ref{stfp}) safely includes 
the region $g_0/f_0\approx -1/3$, which should be the relevant one for 
the normal-to-planar transition in $^3$He.
In Fig.~\ref{rgflowMZM} we also report (thick line) the RG trajectory that 
starts at the O(6) unstable FP and that can be obtained as the limit 
$g_0\to 0^-$ of the trajectories starting at the Gaussian FP. 
Indeed, in this limit the flow first runs close to the $g=0$ axis and
then follows the thick line. This special trajectory is relevant for the 
analysis of the crossover behavior between the O(6) FP and the stable FP 
with $g<0$; see, e.g., Ref.\cite{CPPV-03}. 

It is interesting to note that the estimates of the critical exponents 
(\ref{exp-MZM}) are close to those associated with the chiral FP at
$f^*=-0.3(1)$, $g^*=1.5(1)$, i.e. \cite{PRV-01} $\nu=0.55(3)$ and 
$\eta=0.10(1)$. The reason is that the functions $\beta_{f,g}(f,g)$ and 
$\eta_{\psi,t}(f,g)$ have the approximate symmetry 
\begin{equation}
g\rightarrow \hat{g}=-g, \qquad f\rightarrow \hat{f}=f+g,
\label{apprsym}
\end{equation}
i.e. satisfy 
$\eta_{\psi,t}(f,g) \approx \eta_{\psi,t}(f+g,-g)$, 
$\beta_f(f,g) \approx \beta_f(f+g,-g) + \beta_g(f+g,-g)$, and 
$\beta_g(f,g) \approx - \beta_g(f+g,-g)$. 
Perturbatively, the violations of these relations are much smaller than
the RG functions themselves and appear at 
three loops in the $\beta$-functions and at five loops  
in $\eta_{\psi,t}$.  Moreover, this symmetry is satisfied
by the constant $A(f,g)$, cf. Eq.~(\ref{afg}), that 
controls the large-order behavior. This means that the presence of the 
chiral FP \cite{PRV-01} 
$f^*_{\rm ch} \approx -0.3$, $g^*_{\rm ch} \approx 1.5$, 
implies the presence of another FP for 
$f^* \approx f^*_{\rm ch} + g^*_{\rm ch} \approx 1.2$ and 
$g^* \approx - g^*_{\rm ch} \approx - 1.5$, in substantial agreement with
Eq.~(\ref{stfp}), with approximately the same exponents (\ref{exp-MZM}).
Note that the approximate symmetry (\ref{apprsym}) does not hold 
for $\eta_H(f,g)$ and thus the exponents $\eta_H$, $\beta_H$, and 
$\gamma_H$ differ significantly from the corresponding exponents
at the chiral FP computed in Ref.~\cite{PRV-02}.

In conclusion, the analysis of six-loop series
in the MZM scheme provides a rather robust evidence for a stable FP 
with attraction domain in the region $g_0<0$,
and therefore for the existence of a universality class
associated with the normal-to-planar superfluid transition in $^3$He.
This result contradicts earlier perturbative studies
based on $\epsilon$ expansion \cite{ASV-95} and lower order (three-loop) calculations
within the same scheme \cite{AS-94}, and also a nonperturbative
study of the RG flow of the effective average action
in the lowest order of the derivative expansion \cite{KW-01}.

\section{The massless minimal-subtraction scheme}
\label{sec3}

\begin{figure}[tb]
\centerline{\psfig{width=12truecm,angle=-90,file=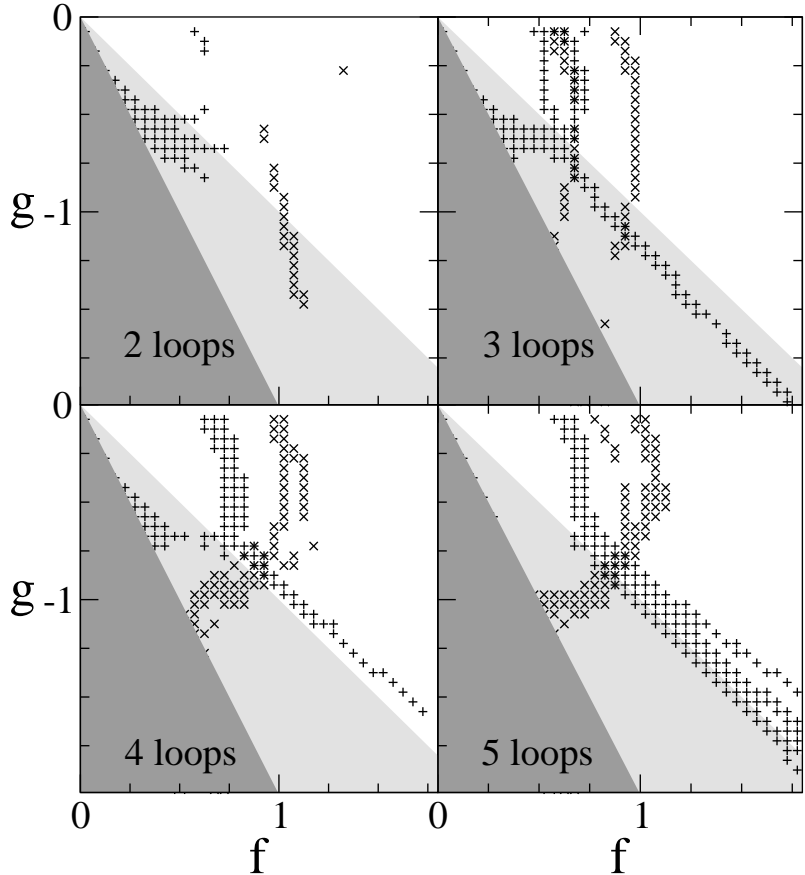}}
\vspace{2mm}
\caption{
Zeroes of the $\beta$-functions in the $f$-$g$ plane in the 
$\overline{\rm MS}$ scheme. Pluses and crosses correspond to zeroes
of $\beta_f$ and $\beta_g$, respectively.
The FP coordinates are: $f^*=0.89(16)$, $g^*=-0.90(24)$.
Different colors are used to mark the regions: (1) $f + g > 0$, 
(2) $f + g < 0$ and $f + {1\over2} g > 0$, (3) $f + {1\over2} g < 0$.
}
\label{zeroesMS}
\end{figure}

In this perturbative approach one considers the massless critical theory 
in dimensional regularization and renormalizes it
in the minimal-subtraction ($\overline{\rm MS}$) scheme \cite{tHV-72}.  
In the standard $\epsilon$-expansion scheme \cite{WF-72}
the FP's, i.e. the common zeroes of the $\beta$-functions,
are determined perturbatively by expanding in powers of $\epsilon$.
Once the expansion of the FP values $f^*$ and $g^*$ is available,
exponents are obtained by expanding
the RG functions at the FP in powers of $\epsilon$.
The $\overline{\rm MS}$ scheme without $\epsilon$ ex\-pan\-sion 
\cite{SD-89} is strictly
related. The RG functions $\beta_{f,g}$ and $\eta_{\psi,t}$ are 
the $\overline{\rm MS}$ functions. However,
$\epsilon$ is no longer considered as a small quantity but it is set to its 
physical value, i.e. in three dimensions one simply sets 
$\epsilon = 1$. Then,  the FP values 
$f^*$, $g^*$  are determined from the common zeroes of the 
resummed $\beta$ functions.
Critical exponents are determined by evaluating
the resummed RG functions $\eta_\psi$ and $\eta_t$ at $f^*$ and $g^*$.
Notice that $f^*$ and $g^*$ have nothing to do with 
the FP values obtained within the MZM scheme, since $f$ and $g$ indicate 
different quantities in the two schemes.

The $\beta$-functions and the RG functions associated with the
standard exponents 
have been computed to five loops in Ref.~\cite{CP-03}
for generic values of $N$. 
Moreover, we computed the five-loop series of the RG dimension
of the operator (\ref{defC}) associated with the magnetic field in $^3$He.
In App.~\ref{appmsb} we report the series for the relevant case $N=3$.

The series are resummed by using the conformal-mapping method.
Semiclassical arguments allow us to compute the large-order behavior of the 
perturbative series. We obtain the same formulae reported
in Sec.~\ref{sec2}, cf. Eqs.~(\ref{lobh}) and (\ref{afg}),
with the only difference that $a=1/2$. 
We should mention that in this approach the series are essentially
four-dimensional. Thus, they may be affected by renormalons 
which make the expansion non-Borel summable for any $f$ and $g$ 
and are not detected by a semiclassical analysis, see, e.g., 
Ref.~\cite{highorder}.
This problem  should also affect the $\overline{\rm MS}$ series of 
the O($N$)-symmetric theories. However, 
the  good agreement between the results of their analysis 
assuming Borel summability \cite{SD-89} and
the estimates  obtained by other methods indicates that renormalon effects are
either very small or absent (note that, as shown in Ref.~\cite{BD-84},
this may occur in some renormalization schemes). 
For example, the analysis \cite{SD-89} of the five-loop $\overline{\rm MS}$
series (using the conformal mapping method and the semiclassical large-order behavior)
gives $\nu=0.629(5)$ for the Ising model and $\nu=0.667(5)$ for the XY model,
which are in good agreement with
the most precise estimates obtained by lattice techniques, 
$\nu=0.63012(16)$ (Ref.~\cite{CPRV-02}) and $\nu=0.63020(12)$ 
(Ref.~\cite{DB-03}) for the Ising model,
and $\nu=0.67155(27)$ (Ref.~\cite{CHPRV-01}) for the XY universality class.
On the basis of these results,
we will assume renormalon effects to be negligible in the analysis of the
two-variable series of the theory at hand.
  
As in the MZM scheme,
two unstable FP's can be easily identified, the Gaussian
FP and the O(6) FP along the $f$-axis.
A stable FP is found in the region $g>0$ \cite{CPPV-03-inprep}, at 
$f^*=0.0(1)$ and $g^*=0.9(1)$, with critical exponents
$\nu=0.61(6)$ and $\eta=0.09(1)$, in reasonable agreement with 
the results of the MZM scheme.

Also in this case there is an approximate symmetry
$g\rightarrow \hat{g}=-g$ and $f\rightarrow \hat{f}=f+g$,
which is violated at three loops in the $\beta$-functions and at five loops   
in the standard critical exponents, and which is satisfied by the large-order behavior.
Thus, we expect a stable FP in the region $g<0$ at
$f^*\approx 0.9$, $g^*\approx  -0.9$,
with exponents approximately equal to those of the chiral FP at $g>0$.

In order to find the zeroes of the $\beta$-functions,
we first resummed the series for the functions
$\beta_f(f,g)/f$ and $\beta_g(f,g)/f$.
For each $\beta$-function we considered several approximants corresponding
to $\alpha$ and $b$ in the range
$0\leq \alpha \leq 4$ and $3\leq b\leq 18$, as in the MZM case.
In Fig.~\ref{zeroesMS} we show the zeroes
of the $\beta$-functions in the region $g<0$.
A common zero is clearly observed at $f^*\approx 0.9$ and $g^*\approx -0.9$.
In order to give an estimate of the FP, we considered all independent 
combinations for the two $\beta$-functions.
Most combinations, approximately  92\%  and 49\% for $l=5$ and 4 respectively, 
have a common zero in the region $g<-0.3$, leading to the estimate
\begin{equation}
f^*=0.89(16), \qquad g^*=-0.90(24),
\label{stfpms}
\end{equation}
where we have again reported two standard deviations as error. 
As before, essentially all zeroes lie in half of the quoted interval---only
a few corresponding to approximants with $\alpha=0$ differ significantly.
Estimates (\ref{stfpms}) are also stable with respect to the number of loops
and in  good agreement with those obtained by using the approximate symmetry.
We also performed a different analysis, in which we resummed 
$B_f(f,g)/f^2$ and $B_g(f,g)/f^2$, cf. Eq.~(\ref{Bdef}), 
obtaining consistent results.  Notice that in this scheme the stable FP is
at the boundary of the region in which perturbative expansions are Borel 
summable, since $f^* + g^* \approx 0$. 
This gives us further confidence on the reliability of the final results.

For the stability eigenvalues,  
we find that in approximately 33\% of the cases the eigenvalues turn out 
to be real, while in
67\% of the cases they are complex. If we average the complex ones we obtain 
$\omega_\pm = 1.2(4+4)\pm i0.8(2+4)$;
real eigenvalues give instead $\omega_1 = 0.9(8+7)$ and 
$\omega_2 = 5(5+1)$. Note that the real results fluctuate wildly, indicating 
that they are probably not reliable. On the other hand, the complex estimate
is more stable and is in good agreement with the MZM result.

Critical exponents are obtained as before.  For the standard exponents we find
\begin{eqnarray}
&& \nu=0.63(3+5),\quad \eta=0.086(13+11),\quad
\gamma=1.20(5+10), \quad \beta = 0.34(2+2),
\label{critMS}
\end{eqnarray}
while for the magnetic (chiral) exponents we find
\begin{equation}
\eta_H=2.63(5+14),\quad
\gamma_H=-0.34(4+7), \quad \beta_H =1.11(4+8).
\label{chexp-MS}
\end{equation}
The error is given as a sum of two terms,
related respectively to the variation with $\alpha$ and $b$ 
(we used $0\leq \alpha\leq 4$ and $3\leq b \leq 18$) and to the uncertainty
of the FP coordinates. Estimates (\ref{critMS}) and  (\ref{chexp-MS}) 
are in good agreement
with the MZM ones: in all cases the difference is significantly smaller than 
the quoted errors, which may be an indication that our error estimates 
are rather conservative. 

\begin{figure}[tb]
\centerline{\psfig{width=12truecm,angle=-90,file=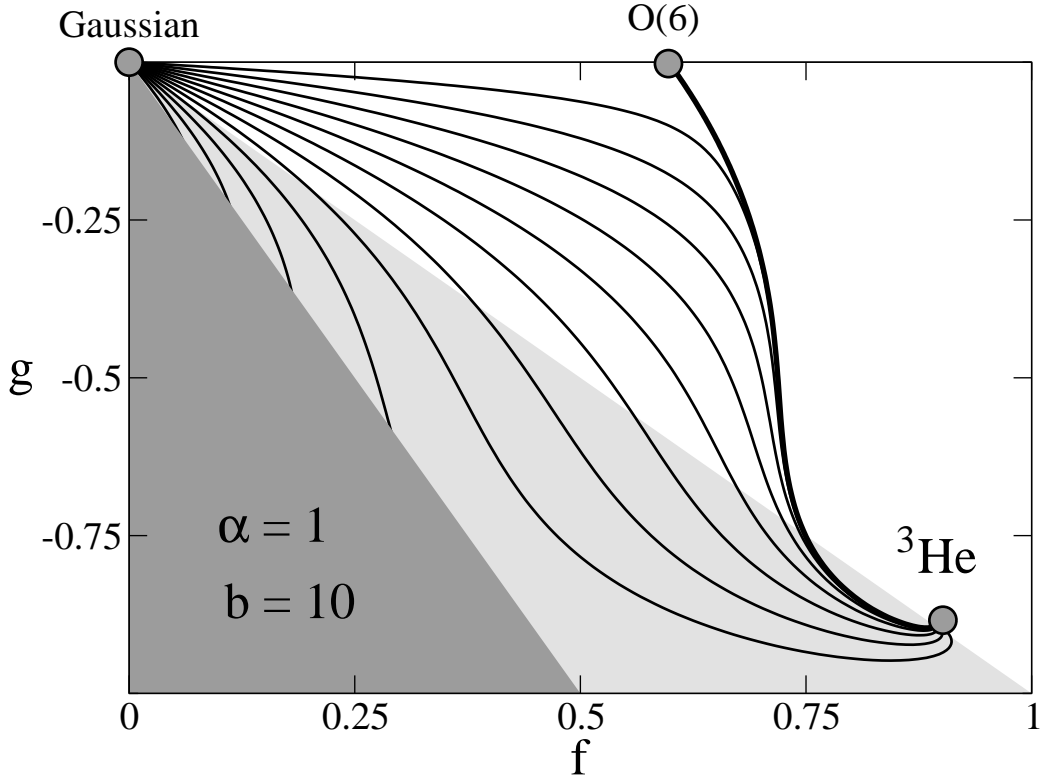}}
\vspace{2mm}
\caption{
RG flow in the $f$-$g$ plane in the $\overline{\rm MS}$ scheme. 
It is obtained by using 
the approximant ($b = 10$, $\alpha = 1$) for $\beta_f(f,g)/f$ and 
$\beta_g(f,g)/f$.
The thick line represents the RG trajectory for $g_0\rightarrow 0^-$.
The other trajectories reaching the FP correspond to (from top to bottom)
$g_0/f_0 = -0.085$, $-0.172$, $-0.261$, $-0.354$, $-0.454$, $-0.561$,
$-0.679$, $-0.810$.
Different colors are used to mark the regions: (1) $f + g > 0$, 
(2) $f + g < 0$ and $f + {1\over2} g > 0$, (3) $f + {1\over2} g < 0$.
}
\label{rgflowMS}
\end{figure}

In Fig.~\ref{rgflowMS} 
we show the RG flow in the quartic-coupling $f,g$ plane for several values of 
$f_0$ and $g_0$ in the region $f_0>0$, $f_0 + g_0>0$, and $g_0<0$.
We report the results for a single approximant for both $\beta$ functions. 
Others give qualitatively similar results.
All trajectories corresponding to $-0.9\lesssim g_0/f_0< 0$
lie in the region $f + {1\over2} g > 0$ and apparently flow towards
the stable FP. Therefore,
the attraction domain of the FP (\ref{stfpms}) safely includes
the region $g_0/f_0\approx -1/3$, which should be relevant for the 
normal-to-planar transition in $^3$He.

\acknowledgments

We thank Pasquale Calabrese and Pietro Parruccini
for useful and interesting discussions. 

\appendix

\section{The six-loop series of the massive zero-momentum scheme}
\label{appgexp}

The theory is renormalized by introducing a set of zero-momentum
renormalization conditions involving the two- and four-point correlation 
functions, i.e. 
\begin{eqnarray}
&&\Gamma^{(2)}_{ai}(p) =
  \delta_{ai} Z_\psi^{-1} \left[ m^2+p^2+O(p^4)\right],
\label{ren1}  \\
&&\Gamma^{(4)}_{abij}(0) =  m Z_\psi^{-2}  {32 \pi\over 9} \left[ 
f {1\over 2} \left( \delta_{ai}\delta_{bj} + \delta_{aj}\delta_{bi} \right)
+ g \delta_{ab}\delta_{ij} \right],
\nonumber
\end{eqnarray}
where $\Gamma^{(2)}$ and  $\Gamma^{(4)}$ are
respectively the Fourier transform of two- and four-point one-particle 
irreducible correlation functions.
In addition, one introduces the function $Z_t(f,g)$ that is defined by the relation
$\Gamma^{(1,2)}_{ai}(0) = \delta_{ai} Z_t^{-1}$,
where $\Gamma^{(1,2)}$ is the one-particle irreducible
two-point function with an insertion of $\psi^*\cdot\psi$.
The FP's of the theory are given by the common zeroes of the
Callan-Symanzik $\beta$-functions
\begin{equation}
\beta_f(f,g) = \left. m{\partial f\over \partial m}\right|_{f_0,g_0},\qquad
\beta_g(f,g) = \left. m{\partial g\over \partial m}\right|_{f_0,g_0}.
\label{betaf}
\end{equation}
The RG functions $\eta_\psi$ and $\eta_t$ associated with
the critical exponents are defined by
\begin{equation}
\eta_{\psi,t}(f,g) = \left. {\partial \ln Z_{\psi,t} \over \partial \ln m}
     \right|_{f_0,g_0}.
\end{equation}
Their relation with the critical exponents is given by
\begin{equation}
\eta = \eta_\psi(f^*,g^*),\qquad
\nu = \left[ 2 + \eta_t(f^*,g^*) - \eta_\psi(f^*,g^*) \right] ^{-1}.
\label{exponents} 
\end{equation}
The RG functions have been computed to 
six loops in Ref.~\cite{PRV-01}.
Here we report the series for $N=3$:
\begin{eqnarray}
\beta_f &=&
-f+
\frac{1}{9}(14\,f^2+8\,f\,g +8\,g^2)
-\frac{16}{2187}(109\,f^3+100\,f^2\,g+177\,f\,g^2+90\,g^3)+
\\
&& +(0.792654\,{f^4} + 1.12206\,{f^3}\,g + 2.63235\,{f^2}\,{g^2} + 2.08529\,f\,{g^3} + 0.127631\,{g^4})+
\nonumber \\
&& -(1.02168\,{f^5} + 1.89478\,{f^4}\,g + 5.18772\,{f^3}\,{g^2} + 5.76831\,{f^2}\,{g^3} + 2.11992\,f\,{g^4}+ 
\nonumber \\
&& + 0.243881\,{g^5})
+(1.56686\,{f^6} + 3.64624\,{f^5}\,g + 11.4687\,{f^4}\,{g^2} + 16.3586\,{f^3}\,{g^3}+
\nonumber \\
&& + 10.9169\,{f^2}\,{g^4} + 3.75387\,f\,{g^5} + 0.480925\,{g^6}) -
 (2.76292\,{f^7} + 7.70253\,{f^6}\,g +
\nonumber \\
&& + 26.9939\,{f^5}\,{g^2} + 46.7321\,{f^4}\,{g^3} + 44.5897\,{f^3}\,{g^4} + 
  25.4728\,{f^2}\,{g^5} + 
\nonumber \\
&& + 7.24362\,f\,{g^6} + 0.637996\,{g^7}),
\nonumber
\end{eqnarray} 

\begin{eqnarray}
\beta_g &=&
-g+
\frac{2}{3}g (2\,f +g)
-\frac{16}{2187}g (127\,f^2+127\,f\,g+6\,g^2)+
\\
&& +g\,(0.926499\,{f^3} + 1.38975\,{f^2}\,g + 0.898551\,f\,{g^2} + 0.216645\,{g^3})+
\nonumber \\
&& -g\,(1.31884\,{f^4} + 2.63769\,{f^3}\,g + 2.87466\,{f^2}\,{g^2} + 1.55675\,f\,{g^3} + 0.179409\,{g^4})+ 
\nonumber \\
&& +g\,(2.10866\,{f^5} + 5.27164\,{f^4}\,g + 8.03232\,{f^3}\,{g^2} + 6.77566\,{f^2}\,{g^3} + 2.29411\,f\,{g^4} +
\nonumber \\
&& + 0.193466\,{g^5}) - g\,(3.9354\,{f^6} + 11.8062\,{f^5}\,g + 22.6695\,{f^4}\,{g^2} + 25.6626\,{f^3}\,{g^3} +
\nonumber \\
&& + 14.855\,{f^2}\,{g^4} + 3.99177\,f\,{g^5} + 0.406903\,{g^6}),
\nonumber
\end{eqnarray}

\begin{eqnarray}
\eta_\psi &=&
\frac{32}{2187}( 2 f^2 + 2 f\,g + 3 g^2)+
\\
&&+(0.00379233\,{f^3} + 0.00568849\,{f^2}\,g + 0.00893906\,f\,{g^2} + 0.00352145\,{g^3})+
\nonumber \\
&&+(0.00873136\,{f^4} + 0.0174627\,{f^3}\,g + 0.0395962\,{f^2}\,{g^2} + 0.0308648\,f\,{g^3} +
\nonumber \\
&&+ 0.00346384\,{g^4}) - (0.0027221\,{f^5} + 0.00680525\,{f^4}\,g + 0.0226066\,{f^3}\,{g^2} +
\nonumber \\
&&+ 0.0271046\,{f^2}\,{g^3} + 0.00857277\,f\,{g^4} + 0.0000120125\,{g^5})+(0.00954181\,{f^6} +
\nonumber \\
&&+0.0286254\,{f^5}\,g + 0.0902133\,{f^4}\,{g^2} + 0.132717\,{f^3}\,{g^3} +0.0958595\,{f^2}\,{g^4} + 
\nonumber \\
&&+0.034245\,f\,{g^5} + 0.00413189\,{g^6}),
\nonumber
\end{eqnarray} 

\begin{eqnarray}
\eta_t &=&
-\frac{1}{9}(8f + 4g)+
\frac{8}{81}( 2f^2 + 2f\,g + 3g^2)+
\\
&&-(0.206554\,{f^3} + 0.309831\,{f^2}\,g + 0.467379\,f\,{g^2} + 0.182051\,{g^3})+
\nonumber \\
&&+ (0.194175\,{f^4} + 0.38835\,{f^3}\,g + 0.930785\,{f^2}\,{g^2} + 0.73661\,f\,{g^3} + 
\nonumber \\
&&+ 0.02992\,{g^4}) - (0.291118\,{f^5} + 0.727795\,{f^4}\,g + 1.88748\,{f^3}\,{g^2} + 
\nonumber \\
&& +2.10343\,{f^2}\,{g^3} + 0.86341\,f\,{g^4} + 0.105499\,{g^5})+(0.444686\,{f^6} + 
\nonumber \\
&&+ 1.33406\,{f^5}\,g + 4.19137\,{f^4}\,{g^2} + 6.15931\,{f^3}\,{g^3} + 4.38435\,{f^2}\,{g^4} + 
\nonumber \\
&&+ 1.527\,f\,{g^5} + 0.176483\,{g^6}).
\nonumber
\end{eqnarray} 
The critical exponents associated with a real magnetic external field 
are determined from the RG dimension 
of the operator $C_{a}$, cf. Eq.~(\ref{defC}). In order to compute it,
we define the renormalization function $Z_c(f,g)$ from the one-particle 
irreducible two-point function $\Gamma^{(c,2)}$ with an insertion of the 
operator $C_{a}$, i.e.
\begin{equation}
\Gamma^{(c,2)}(0)_{aij} = Z_c^{-1} \; T_{aij},
\end{equation}
where $T_{aij}$ is an appropriate tensor normalized so that $Z_c=1$
at tree level. The RG function 
\begin{equation}
\eta_{c}(f,g) = \left. {\partial \ln Z_{c} \over \partial \ln m} \right|_{f_0,g_0}
\end{equation}
was computed to six-loop in Ref.~\cite{PRV-02}.
For the three-component case one obtains 
\begin{eqnarray}
\eta_c &=& -{2\over9} (f - 2 g) + {8\over 243} (3 {f^2} - 2 \,f\,g - 3\,{g^2})+
\nonumber \\
&&-(0.0426744\,{f^3} - 0.108535\,{f^2}\,g - 0.0512203\,f\,{g^2} + 0.0777653\,{g^3})
\nonumber \\
&&+(0.0646127\,{f^4} - 0.0340279\,{f^3}\,g - 0.00236481\,{f^2}\,{g^2} + 0.0275227\,f\,{g^3}+ 
\nonumber \\
&& - 0.0251600\,{g^4} )-(0.0713318\,{f^5} - 0.0868275\,{f^4}\,g - 0.0504380\,{f^3}\,{g^2}+ 
\nonumber \\
&& + 0.00425084\,{f^2}\,{g^3} - 0.0721756\,f\,{g^4} - 
  0.0120727\,{g^5})+(0.125344\,{f^6}+
\nonumber \\
&&  - 0.0394498\,{f^5}\,g + 0.0763944\,{f^4}\,{g^2} + 0.0550844\,{f^3}\,{g^3} - 0.194565\,{f^2}\,{g^4}+
\nonumber \\
&&-  0.0632578\,f\,{g^5} + 0.0152157\,{g^6}).
\nonumber
\end{eqnarray} 
The critical exponents $\eta_H$, $\gamma_H$, and $\beta_H$ are obtained by
\begin{equation}
\eta_H= 1 - 2 \eta_c + 2 \eta,
\quad 
\gamma_H = \nu (2 - \eta_H),
\quad
\beta_H= \nu ( 1 - \eta_c + \eta),
\label{hrel}
\end{equation}
where $\eta_c\equiv \eta_c(f^*,g^*)$.

\section{The five-loop series in the minimal-subtraction scheme}
\label{appmsb}

In the $\overline{\rm MS}$ scheme one sets
\begin{eqnarray}
\psi &=& [Z_\psi(f,g)]^{1/2} \psi_R, \\
f_0 &=& A_d \mu^\epsilon F(f,g) , \nonumber \\
g_0 &=& A_d \mu^\epsilon G(f,g) , \nonumber
\end{eqnarray}
where the renormalization functions $Z_\psi$, $F$, and $G$
are determined from the divergent part of the two- and four-point
correlation functions computed in dimensional regularization.
They are normalized  so that
$Z_\psi(f,g) \approx 1$, $F(f,g) \approx f$, and 
$G(f,g) \approx g$ at tree level. 
Here $A_d$ is a $d$-dependent constant given 
by $A_d= 2^d \pi^{d/2} \Gamma(d/2)/3$. 
We also introduce a mass renormalization constant $Z_t(f,g)$ by requiring 
$Z_t \Gamma^{(1,2)}$ to be finite when expressed in terms of $f$ and $g$. 
Here $\Gamma^{(1,2)}$ is the two-point function with an insertion of 
$\psi^*\cdot \psi$.
Once the renormalization constants are determined, one computes 
the $\beta$ functions from 
\begin{equation}
\beta_f (f,g) = \mu \left. {\partial f \over \partial \mu} \right|_{f_0,g_0},
\qquad\qquad
\beta_g (f,g) = \mu \left. {\partial g \over \partial \mu} \right|_{f_0,g_0},
\end{equation}
and the critical RG functions $\eta_\psi$ and $\eta_t$ from
\begin{equation}
\eta_{\psi,t}(f,g) 
=  \left. {\partial \log Z_{\psi,t} \over \partial \log \mu} \right|_{f_0,g_0}.
\end{equation}
Their relation with the critical exponents is given in Eq.~(\ref{exponents}).
The $\beta$-functions have a simple dependence on $d$: 
\begin{equation}
\beta_f(f,g) = (d-4) f + B_f(f,g),\qquad
\beta_g(f,g) = (d-4) g + B_g(f,g),
\label{Bdef}
\end{equation}
where the functions $B_f(f,g)$ and $B_g(f,g)$---as well as 
the RG functions $\eta_{\psi,t}(f,g)$---are independent of $d$.

In the following we report the five-loop series \cite{CP-03} for $N=3$:
\begin{eqnarray}
B_f &=&
\frac{1}{3}(7\,f^2+4\,f\,g +4\,g^2)
-\frac{1}{9}(24\,f^3+22\,f^2\,g+39\,f\,g^2+20\,g^3)+
\\
&& +(9.07446\,{f^4} + 12.5703\,{f^3}\,g + 29.4275\,{f^2}\,{g^2} + 23.8424\,f\,{g^3} + 2.41477\,{g^4})+
\nonumber \\
&& -(46.7683\,{f^5} + 85.6528\,{f^4}\,g + 227.316\,{f^3}\,{g^2} + 256.217\,{f^2}\,{g^3} + 110.777\,f\,{g^4} +
\nonumber \\
&& + 16.3103\,{g^5})+(296.166\,{f^6} + 677.379\,{f^5}\,g + 2026.83\,{f^4}\,{g^2} + 2898.93\,{f^3}\,{g^3} +
\nonumber \\
&& + 2096.65\,{f^2}\,{g^4} + 791.170\,f\,{g^5} + 109.827\,{g^6}),
\nonumber \\
B_g &=&
g\, [\,(2\,f+g)
-\frac{1}{9}(28\,f^2+28\,f\,g+g^2)+(11.1573\,{f^3} + 16.7359\,{f^2}\,g+
\\
&&  + 9.75736\,f\,{g^2} + 2.06019\,{g^3})-(62.5357\,{f^4} + 125.071\,{f^3}\,{g} + 123.278\,{f^2}\,{g^2} +
\nonumber \\
&&  +60.7252\,f\,{g^3} + 10.5496\,{g^4})+(422.235\,{f^5} + 1055.59\,{f^4}\,g + 1458.99\,{f^3}\,{g^2} + 
\nonumber \\
&&  +1132.71\,{f^2}\,{g^3} + 411.714\,f\,{g^4} + 52.1814\,{g^5})\, ],
\nonumber 
\end{eqnarray}

\begin{eqnarray}
\eta_\psi &=&\frac{1}{18}(2\,f^2+2\,f\,g+3\,g^2)-\frac{1}{216}(14\,{f^3} + 21\,{f^2}\,g + 33\,f\,{g^2} + 13\,{g^3})+
\\
&&+\frac{5}{2592}( 86\,{f^4} + 172\,{f^3}\,g + 474\,{f^2}\,{g^2} + 388\,f\,{g^3} - 9\,{g^4})+
\nonumber \\
&&-(0.709208\,{f^5} + 1.77302\,{f^4}\,g + 4.78026\,{f^3}\,{g^2} + 5.39737\,{f^2}\,{g^3} + 
\nonumber \\
&& 2.31434\,f\,{g^4} + 0.325646\,{g^5}),
\nonumber \\
\eta_t &=&-\frac{2}{3}(2\,f+g)+\frac{1}{3}(2\,f^2+2\,f\,g+3\,g^2)+
\\
&&-\frac{1}{108}(208\,{f^3} + 312\,{f^2}\,g + 486\,f\,{g^2} + 191\,{g^3})+(7.00217\,{f^4} + 14.0043\,{f^3}\,g +
\nonumber \\
&& +29.2911\,{f^2}\,{g^2} + 22.2890\,f\,{g^3} + 3.64149\,{g^4})-
    (33.6900\,{f^5} + 84.2251\,{f^4}\,g + 
\nonumber \\
&&+207.371\,{f^3}\,{g^2} + 226.832\,{f^2}\,{g^3} + 
108.339\,f\,{g^4} + 19.1618\,{g^5}).
\nonumber
\end{eqnarray}
We also computed the $\overline{\rm MS}$ expansion of the RG dimension
of the operator $C_{a}$, cf. Eq.~(\ref{defC}). For this purpose,
we computed the renormalization constant $Z_c(f,g)$ by requiring
$Z_c \Gamma^{(c,2)}$ to be finite when expressed in terms of $f$ and $g$, 
where $\Gamma^{(c,2)}$ is the one-particle irreducible
two-point function with an insertion of the operator
$C_{a}$. Then, one defines the corresponding RG function
\begin{equation}
\eta_{c}(f,g) 
=  \left. {\partial \log Z_{c} \over \partial \log \mu} \right|_{f_0,g_0}.
\end{equation}
The resulting series for the three-component case is given by
\begin{eqnarray}
\eta_c &=&-\frac{1}{3}(f-2\,g)+\frac{1}{9}(3\,f^2-2\,f\,g-3\,g^2)-\frac{5}{216}(22\,{f^3} - 42\,{f^2}\,g - 27\,f\,{g^2} + 26\,{g^3})+
\nonumber \\
&& +(1.93436\,{f^4} - 2.50866\,{f^3}\,g - 2.13390\,{f^2}\,{g^2} + 0.910908\,f\,{g^3} + 0.285515\,{g^4})+	 
\nonumber \\
&&-(8.73201\,{f^5} - 9.61852\,{f^4}\,g - 9.00230\,{f^3}\,{g^2} + 1.92888\,{f^2}\,{g^3} - 0.333945\,f\,{g^4}+
\nonumber \\
&& - 0.810730\,g^5).
\label{etac-pertMS}
\end{eqnarray}
The critical exponents $\eta_H$, $\gamma_H$ and $\beta_H$ are computed by
using  relations (\ref{hrel}). The expansion of $\eta_c$ was computed for 
generic values of $N$. 
We verified that $\eta_c(f,0)$ coincides with the expansion of 
the RG function $\eta_T$, related to the insertion of 
the spin-2 operator in the two-point function, in
the O($2 N$) theory obtained by setting $g=0$ (perturbative five-loop series 
for $\phi_T = (2 + \eta_T - \eta)\nu$ 
were computed in Ref.~\cite{CPV-03}).
Moreover, for $N=2$ one can prove and check the relation
$\eta_c(2x/3,-x/3)= \eta_t(2x/3,-x/3)$, and 
$\eta_{t,\phi}(2x/3,-x/3)=\eta^{XY}_{t,\phi}$
where $\eta^{XY}_{t,\phi}$ are the RG functions of the XY 
(O(2)-symmetric) theory,
which can be found in Ref.~\cite{KNSCL-93}.



\begin{references}
\bibitem{footnoteH}
Alternatively, one may write the Hamiltonian
in terms of two real three-component fields
$\phi_a$ ($a=1,2$) as
\begin{eqnarray}
&&{\cal H}  = \int d^d x
 \left\{ {1\over2}
      \sum_{a} \Bigl[ (\partial_\mu \phi_{a})^2 + r \phi_{a}^2 \Bigr]
+ {1\over 4!}u_0 \Bigl( \sum_a \phi_a^2\Bigr)^2
+ {1\over 4!}  v_0
\sum_{a,b} \Bigl[ ( \phi_a \cdot \phi_b)^2 - \phi_a^2\phi_b^2\Bigr]
             \right\},
\nonumber
\end{eqnarray}
where $u_0=\case{3}{2} (f_0+g_0)$ and $v_0=3g_0$.
\bibitem{footH}
In terms of the order parameter $A_{ai}$ that controls the transition in
the absence of dipolar interactions, the magnetic field $\vec{H}$ 
couples $A_{ai}$ as \cite{MS-73}
\[
i\alpha \sum_{abc} H_a (AA^\dagger)_{bc} \epsilon_{abc} 
+ \beta \sum_{ab} H_a H_b (AA^\dagger)_{ab}.
\]
In the presence of dipolar interactions $A_{ai} = \epsilon_{aib}\psi_b$, so 
that we obtain 
\[
i\alpha \sum_{abc} H_a \psi^*_b \psi_c \epsilon_{abc} + 
\beta \sum_{ab} H_a H_b (\psi^*_a \psi_b - \delta_{ab} |\psi|^2).
\]
An additional term is also possible, $H^2 {\rm Tr}\, AA^\dagger$, which
is found however to be negligibly small (Ref.~\cite{MS-73}).
Of course, for $H\to 0$, only the linear term is relevant.
\end{references}
\end{document}